\documentstyle[psfig]{mn}

\title[Radial velocity measurements of white dwarfs]
{Radial velocity measurements of white dwarfs
}
\author[P. F. L. Maxted ,  T.~R.~Marsh, C.~K.~J.~Moran]
       {P.~F.~L.~Maxted,  T.~R.~Marsh, C.~K.~J.~Moran, \\ 
        University of Southampton, Department of Physics \& Astronomy,
        Highfield, Southampton, S017 1BJ, UK}
\date{Accepted 1999 
      Received 1999 }

\pagerange{\pageref{firstpage}--\pageref{lastpage}}
\pubyear{1999}

\newcommand{\Msolar}{\mbox{${\rm M}_{\odot}\,$}}

\begin{document}

\maketitle

\label{firstpage}

\begin{abstract}
 We present 594 radial velocity measurements for 71 white dwarfs obtained
during our search for binary white dwarfs and not reported elsewhere. We
identify three excellent candidate binaries, which require further
observations to confirm our preliminary estimates for their orbital periods,
and one other good candidate. We investigate whether our data support the
existence of a population of single, low mass ($\la 0.5\Msolar$) white dwarfs
(LMWDs). These stars are difficult to explain in standard models of stellar
evolution. We find that a model with a mixed single/binary population is at
least $\sim 20 $ times more likely to explain our data than a pure binary
population. This result depends on assumed period distributions for binary
LMWDs, assumed companion masses and several other factors. Therefore, the
evidence in favour of the existence of a population of single LMWDs is not
sufficient, in our opinion, to firmly establish the existence of such a
population, but does suggest that extended observations of LMWDs to obtain a
more convincing result would be worthwhile .

\end{abstract} 

\section{Introduction}
 The observed mass distribution of white dwarf stars is strongly peaked
around 0.55\Msolar (Finley et~al. 1997, Bergeron et~al. 1992, Bragaglia
et~al. 1995). Although models of the evolution leading to white dwarfs are
extremely uncertain, it appears that this is the minimum mass of a white dwarf
that can be formed through single star evolution in the lifetime of the
Galaxy (Bragaglia et~al. 1995). White dwarfs more massive than this minimum
are formed from initially more massive stars, but they are much less common
than lower mass stars, and so the observed mass distribution is strongly
peaked. In this paper we deal with white dwarf stars below this ``minimum''
mass. These are thought to be the result of binary star evolution, in which
the evolution of a star during the red giant phase is interrupted by
interactions with a nearby  star. The physics of this interaction is complex
but it is thought to lead to the stripping of the outer hydrogen layers from
the red giant in a ``common-envelope'' phase, halting the formation of the
degenerate helium core and leading to the formation of an anomalously low mass
white dwarf (Iben \& Livio 1993). The hypothesis that binary evolution forms
low mass white dwarfs (LMWDs) was confirmed by the discovery of Marsh et~al.
(1995) of at least 5 short period binary white dwarfs in a sample of 7 LMWDs.
However, there is growing evidence that LMWDs may not all be binaries (Maxted
\& Marsh 1998). It has been suggested that this is a result of the merging of
the binary following the common-envelope phase (Iben et~al. 1997), but the
lack of any detectable rotation in the apparently single white dwarfs has cast
doubt on this suggestion (Maxted \& Marsh 1998). An alternative hypothesis is
that the giant planets recently discovered orbiting solar-type stars lead to a
common-envelope phase, but evaporate during that phase leaving an apparently
single LMWD (Nelemans \& Tauris 1998).

 We have been successful in finding new white dwarf binaries and measuring
their orbital periods using the techniques of Marsh et~al. (1995). Those
results have been presented elsewhere (Moran 1999; Moran, Marsh \& Maxted 2000;
 Maxted , Marsh \& Moran, 2000). We have observed many white dwarfs
in the course of our  search for binary white dwarfs but have not, in general,
reported these radial velocity measurements unless the star was found to be a
binary and the orbital period identified. These radial velocity measurements
are a valuable resource, both for kinematic studies and for future surveys for
binary white dwarfs. Therefore, in this paper we report our 594 radial
velocity measurements for 71 white dwarfs not already reported elsewhere. We
identify 4 new candidate binary white dwarfs and report preliminary orbital
periods for three of them. We also consider the evidence for the existence of
a population of single low mass white dwarfs.

\section{Observations and reductions}
 The data have been obtained over several years using several
instruments.  Most of the data come from observations obtained with the
intermediate dispersion spectrograph (IDS) on the 2.5m Isaac Newton Telescope
(INT) on the Island of La Palma. Additional spectra for some stars  were
obtained using the ISIS spectrograph on the 4.2m William Herschel Telescope
(WHT), also on La Palma,  and the RGO spectrograph on the 3.9m
Anglo-Australian Telescope (AAT) at Siding Spring, Australia. The detectors
used in every case were charge-coupled devices (CCDs).   Details of all three
instruments, the dates of all the observing runs and the dispersion per
pixel used
are given in Table~\ref{ObsTable}. 

The observing procedure is very similar in each case. We obtain spectra of our
target stars around the H$\alpha$ line with a resolution of $\la $1\AA.
 Exposure times are typically 5--20\,minutes and never longer than 30
minutes. Spectra of an arc lamp are taken before and after each target
spectrum with the telescope tracking the star. None of the CCDs used showed
any structure in unexposed images, so a constant bias level determined from a
clipped-mean value in the over-scan region was subtracted from all the images.
Sensitivity variations were removed using observations of a tungsten
calibration lamp. The sensitivity variations along the spectrograph slit are
removed using observations of the twilight sky in the AAT images because the
tungsten calibration lamp is inside the spectrograph. We have occasionally
used the same technique for the WHT and INT spectra, though it makes little
difference in practice whether we use sky images or lamp images to calibrate
these images.

Extraction of the spectra from the images was performed automatically using
optimal extraction to maximize the signal-to-noise of the resulting spectra
(Horne 1986).  The arcs associated with each stellar spectrum were extracted
using the profile determined for the stellar image to avoid possible
systematic errors due to tilted arc lines. The wavelength scale was determined
from a polynomial fit to measured arc line positions and the wavelength of the
target spectra interpolated from the calibration established from the
bracketing arc spectra. Uncertainties on every data point calculated from
photon statistics are rigorously propagated through every stage of the data
reduction.

\begin{table}
\caption{\label{ObsTable} Summary of the spectrograph/telescope combinations
used to obtain spectra for this study. The slit width used in each case is
approximately 1arcsec. The resolution and the sampling are both in units of
\AA.}
\begin{tabular}{@{}lllrr} 
Date &Telescope & Spectro-&  Reso-& Sampling \\
     &          & graph   & lution & \\
Mar 96 & AAT & RGO &  0.7  & 0.23 \\
Aug 97 & AAT & RGO &  0.7  & 0.24 \\
Mar 97 & AAT & RGO &  0.7  & 0.23 \\
Jun 98 & AAT & RGO &  0.7  & 0.29 \\
Mar 99 & AAT & RGO &  0.7  & 0.29 \\
Apr 94 & INT & IDS & 0.7  & 0.36 \\
Jun 95 & INT & IDS & 0.9  & 0.39 \\
Feb 97 & INT & IDS & 0.9  & 0.39 \\
Jun 97 & INT & IDS & 0.9  & 0.39 \\
Nov 97 & INT & IDS & 0.9  & 0.39 \\
Feb 98 & INT & IDS & 0.9  & 0.39 \\
Sep 98 & INT & IDS & 0.9  & 0.39 \\
Feb 99 & INT & IDS & 0.9  & 0.39 \\
Apr 99 & INT & IDS & 0.6  & 0.30 \\
Jun 93 & WHT & ISIS & 0.8 & 0.38 \\
Aug 93 & WHT & ISIS & 0.8 & 0.38 \\
Jul 94 & WHT & ISIS & 1.8 & 0.74 \\
Jan 95 & WHT & ISIS & 0.8 & 0.40 \\
Nov 97 & WHT & ISIS & 0.8 & 0.40 \\
Feb 98 & WHT & ISIS & 0.8 & 0.40 \\
Jul 98 & WHT & ISIS & 0.8 & 0.40 \\
\end{tabular}
\end{table}

\begin{table}
\caption{\label{OffsetTable} Measurements of offsets in radial velocity
measurements between various data sets.}
\begin{tabular}{@{}lllr}
Star & First         & Second        & Offset \\
     & observing run & observing run & (km\,s$^{-1}$) \\
WD\,0132+254   & WHT, Nov 97  & INT, Feb 98 & $  +6.9 \pm 4.0$ \\
WD\,0316+345   & WHT, Jan 95  & INT, Sep 98 & $  -0.5 \pm 1.6$ \\
WD\,0341+021   & WHT, Nov 97  & INT, Feb 98 & $ -31.5 \pm 6.9$ \\
WD\,0401+250   & INT, Nov 97  & INT, Feb 98 & $  +1.6 \pm 2.7$ \\
WD\,0401+250   & WHT, Nov 97  & INT, Feb 98 & $  -1.9 \pm 2.7$ \\
WD\,0437+152   & WHT, Nov 97  & INT, Feb 98 & $  -0.9 \pm 3.0$ \\
WD\,0453+418   & WHT, Jan 95  & INT, Feb 99 & $  -3.4 \pm 1.2$ \\
WD\,0549+158   & INT, Feb 98  & INT, Feb 99 & $  -1.9 \pm 3.0$ \\
WD\,0808+595   & WHT, Nov 97  & INT, Feb 98 & $  -2.6 \pm 6.5$ \\
WD\,1031$-$114 & INT, Feb 99  & AAT, Mar 99 & $  -3.4 \pm 2.7$ \\
WD\,1105$-$048 & AAT, Mar 97  & INT, Feb 99 & $  +0.3 \pm 1.2$ \\
WD\,1105$-$048 & AAT, Mar 96  & INT, Feb 99 & $  +0.5 \pm 0.7$ \\
WD\,1257+032   & WHT, Jan 95  & INT, Jun 95 & $  +3.2 \pm 2.5$ \\
WD\,1257+032   & INT, Jun 95  & AAT, Mar 99 & $  -0.9 \pm 5.6$ \\
WD\,1257+032   & WHT, Jan 95  & AAT, Mar 99 & $  +2.3 \pm 5.7$ \\ 
WD\,1310+583   & INT, Jul 97  & INT, Feb 98 & $  +1.2 \pm 2.2$ \\
WD\,1327$-$083   & AAT, Mar 97  & AAT, Jun 98 & $  -1.3 \pm 0.3$ \\
WD\,1327$-$083   & AAT, Mar 96  & AAT, Jun 98 & $  +1.1 \pm 0.2$ \\ 
WD\,1353+409   & WHT, Jun 93  & WHT, Jan 95 & $  -3.8 \pm 3.5$ \\
WD\,1353+409   & WHT, Jan 95  & WHT, Feb 98 & $  +9.3 \pm 3.2$ \\
WD\,1407$-$475   & AAT, Mar 96  & AAT, Mar 97 & $ -22.0 \pm 1.5$ \\
WD\,1614+136   & WHT, Jun 93  & WHT, Feb 98 & $  +2.1 \pm 2.1$ \\
WD\,1614+136   & WHT, Jun 93  & AAT, Jun 96 & $  +1.9 \pm 4.7$ \\
WD\,1620$-$391   & AAT, Mar 97  & AAT, Aug 97 & $  +3.6 \pm 0.6$ \\
WD\,1620$-$391   & AAT, Mar 97  & AAT, Jun 98 & $  -0.8 \pm 1.6$ \\
HS\,1653+7753  & INT, Sep 98  & INT, Feb 99 & $  -3.0 \pm 8.2$ \\
WD\,1943+163   & INT, Jun 95  & AAT, Jun 95 & $  +0.8 \pm 2.1$ \\
\end{tabular}
\end{table}

\section{Analysis}
\subsection{Radial velocity measurements.}
 To measure the radial velocities we used least-squares fitting of a
model line profile. This model line profile is the summation of four Gaussian
profiles with different widths and depths but with a common central position
which varies between spectra.  Only data within 5000\,km\,s$^{-1}$ of the
H$\alpha$ line is included in the fitting process. We first normalize the
spectra using a linear fit to the continuum either side of the H$\alpha$ line.
We then use a least-squares fit to all the spectra to establish the shape of
the model line profile. A least squares fit of this profile to each spectrum
in which the position of the line is the only free parameter gives  the final
heliocentric radial velocities reported in Table~\ref{RVTable}. The
uncertainties quoted are calculated by propagating the uncertainties on every
data point in the spectra right through the data reduction and analysis. These
uncertainties are reliable in most cases, but some caution must be exercised
for quoted uncertainties of less than $\sim 0.5$\,km\,s$^{-1}$. This
corresponds to less than $1/20$ of a pixel in the original data, so systematic
errors such as telluric absorption features and uncertainties in the
wavelength calibration are certain to be a significant source of uncertainty
for these measurements. 

 Where data has been obtained for a star on more than one instrument we have
measured the offset between the data sets to look for systematic differences.
These offsets are given in Table~\ref{OffsetTable}. Almost all of these
offsets are consisted with an offset between data sets of no more than $\sim
1$\,km\,s$^{-1}$. The obvious exceptions are  WD\,0341+021 and WD\,1407$-$475,
which we discuss more fully below.

\subsection{Criterion for variability.}
 For each star we calculate a weighted mean radial velocity. This mean is the
best estimate of the radial velocity of the star assuming this quantity is
constant. We then calculate the $\chi^2$ statistic for this ``model'', i.e.
the goodness-of-fit of a constant to the observed radial velocities. We can
then compare the observed value of $\chi^2$ with the distribution of $\chi^2$
for the appropriate number of degrees of freedom. We then calculate the
probability of obtaining the observed value of $\chi^2$ or higher from random
fluctuations of constant value, $p$. The observed values of the weighted mean
radial velocity, $\chi^2$ and the logarithm of this probability,
$\log_{10}(p)$, are given for all the white dwarfs in our sample in
Table~\ref{ResultsTable}. If we find $\log_{10}(p) < -4$ we consider this to
be a detection of a binary. In a sample of 71 objects, this results in a less
than 1\,percent chance of random fluctuations producing one or more false
detections.

 In order to estimate the fraction of binaries that would be detected using
our observations with this detection criterion we use a Monte Carlo approach.
We generate synthetic radial velocity measurements with the same temporal
sampling and accuracy as the actual observations of each star and add the
appropriate amount of noise. We include the projection effects due to randomly
oriented orbits. Periods are selected randomly from one of the theoretical
period distributions described below. The mass of the  white dwarf observed,
$M$, is taken from Table~\ref{ResultsTable} if known or is calculated for each
trial perid $P$ from  $\log(M) = 0.13 \log(P) -0.6$. This is simply an
approximation to the main feature of the bivariate distribution of periods and
masses for binary white dwarfs given by Saffer et~al. (1998). We then estimate
our detection efficiency using the number of trials which statisfy our
detection criterion for the following two cases.

The first case is a white dwarf companion with the same mass as the visible
white dwarf. We use the sum of the period distributions for white dwarfs with
white dwarf companions of all types including the loss of systems due to
$10^8$y of gravitational wave radiation given by Iben et~al. (1997, their
Figs~2(c) and 2(d)). Note that their models give a mean mass ratio of around
0.7 with the fainter, i.e., older, companion being more massive. However,
there are now six white dwarf -- white dwarf binaries with directly measured
mass ratios, and these tend to be $\ga 1$ (Table~\ref{QTable}). It is not
straightforward to estimate the selection effects but a mass ratio of 1 does
seem to be more typical for these binaries. This detection efficiency is
given in Table~\ref{ResultsTable} under $e(A)$. 

\begin{table}
\caption{\label{QTable} Measured mass ratios for white dwarf -- white dwarf
binaries.}
\begin{tabular}{@{}lrl}
Name & Mass ratio & Reference \\
WD\,0136+768 & 1.27$\pm$0.04 &Moran, Maxted \& Marsh 2000 \\
WD\,0135-052 & 0.90$\pm$0.04 &Saffer et~al. 1988 \\
WD\,0957$-$666 & 1.14$\pm$0.02 &Moran, Maxted \& Marsh 2000 \\
WD\,1101+364 & 0.87$\pm$0.03 &Marsh 1995 \\
WD\,1204+450 & 1.095$\pm$0.04 &Moran, Maxted \& Marsh 2000 \\
WD\,1704+481.2 & 0.70$\pm$0.03 &Maxted, Marsh \& Moran 2000 \\
\end{tabular}
\end{table}

\begin{figure}
\leavevmode\centering{
\psfig{file=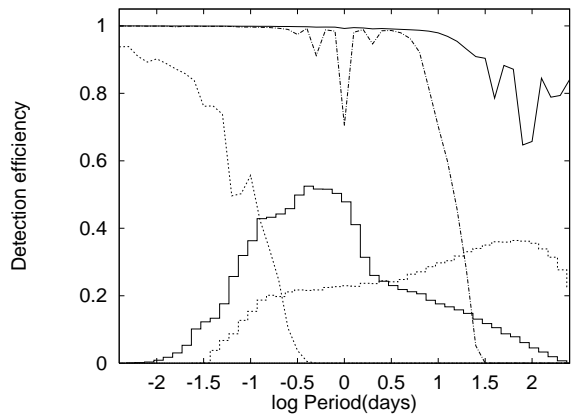,width=0.45\textwidth} }
\caption{\label{DetectionEfficiencyFig} The detection efficiency as a function
of orbital period assuming a mass ratio of one for  WD\,0346$-$011 (dashed
line), WD\,0913+442, (dash-dotted line) and WD\,0101+048 (solid line). The
theoretical period distributions used to estimate the detection efficiencies
for stars with  white dwarf companions (histogram, solid line) and main
sequence companions (histogram, dashed line) are also shown.} 
\end{figure}

 The second case is a main-sequence companion with a mass of 0.08\Msolar. The
theoretical period distribution in this case is the sum of the distributions
given by Iben et~al. for companions to white dwarfs with mass less than
0.3\Msolar (their Figs~3(c) and 3(d)). This detection efficiency is given in
Table~\ref{ResultsTable} under $e$(B). We use 100,000 trials to measure these
efficiencies, which is sufficient to give an accuracy of a few tenths of one
percent. We can also plot these detection efficiencies as a function of period
to get a more qualitative view. Some examples are shown in
Fig.~\ref{DetectionEfficiencyFig}.

\begin{table*}
\caption{\label{ResultsTable}
 Summary of our radial velocity measurements for white dwarfs. 
 References for the masses are as follows: 
1.~Bergeron et~al. 1992;
2.~Bergeron et~al. 1995;
3.~Finley et~al. 1997;
4.~Homeier et~al. 1998;
5.~Moran 1999;
6.~Vennes et~al. 1997;
7.~Bragaglia et~al.1995.
 }
\begin{tabular}{@{}lrrrrrrlr}
Name & \multicolumn{1}{l}{N}&\multicolumn{1}{l}{Mean} & 
\multicolumn{1}{l}{$\chi^2$} & $\log_{10}(p)$ & \multicolumn{1}{l}{$e(A)$}& 
\multicolumn{1}{l}{$e$(B)} & Mass &\multicolumn{1}{l}{Ref.} \\
 & & \multicolumn{1}{l}{(km\,s$^{-1}$)}  & & & \multicolumn{1}{l}{(\%)} &
\multicolumn{1}{l}{(\%)} & (\Msolar) \\
WD\,0011+000 & 3 & 27.0 $\pm$ 2.3 & 2.80 & -0.61& 66.6 & 21.5 &  \\
WD\,0101+048 & 14 & 63.4 $\pm$ 0.2 &105.27 &-15.80& 99.1 & 68.2 &  \\
WD\,0126+101 & 6 & 7.1 $\pm$ 0.4 & 5.90 & -0.50& 90.8 & 40.8 & 0.50 $\pm$0.03 & 5 \\
WD\,0132+254 & 8 & 36.3 $\pm$ 0.5 & 5.71 & -0.24& 95.5 & 49.1 & 0.36 $\pm$0.03 & 5 \\
WD\,0142+312 & 8 & 37.6 $\pm$ 1.1 & 11.41 & -0.92& 86.1 & 35.7 &  \\
WD\,0143+216 & 5 & 20.8 $\pm$ 1.2 & 7.17 & -0.89& 82.5 & 32.2 &  \\
WD\,0147+674 & 6 & 30.2 $\pm$ 1.1 & 9.65 & -1.07& 79.9 & 22.6 & 0.45$\pm$0.03,0.48$\pm$0.01 & 1,3 \\
WD\,0148+467 & 2 & 15.3 $\pm$ 2.4 & 0.44 & -0.29& 34.3 & 4.3 & 0.53$\pm$0.03,0.57$\pm$0.03 & 1,5 \\
WD\,0151+017 & 4 & 63.2 $\pm$ 0.8 & 0.64 & -0.05& 72.1 & 21.5 & 0.48 $\pm$0.03 & 5 \\
WD\,0213+396 & 6 & 26.2 $\pm$ 1.4 & 22.36 & -3.35& 86.5 & 36.8 &  \\
WD\,0316+345 & 12 & -42.9 $\pm$ 0.2 & 27.82 & -2.46& 98.4 & 64.8 & 0.40$\pm$0.03 & 1 \\
WD\,0320$-$539 & 7 & 57.8 $\pm$ 0.8 & 14.21 & -1.56& 87.8 & 30.0 & 0.58$\pm$0.02,0.47$\pm$0.03 & 3,7 \\
WD\,0332+320 & 4 & 100.9 $\pm$ 1.7 & 2.60 & -0.34& 66.2 & 11.8 & 0.71$\pm$0.03 & 5 \\
WD\,0339+523 & 9 & 3.3 $\pm$ 0.5 & 7.41 & -0.31& 92.8 & 46.5 & 0.34$\pm$0.03 & 1 \\
WD\,0341+021 & 7 & -53.2 $\pm$ 0.6 & 33.65 & -5.11& 94.1 & 45.6 & 0.38$\pm$0.03 & 5 \\
WD\,0346$-$011 & 10 & 134.5 $\pm$ 5.2 & 26.95 & -2.85& 55.2 & 0.0 & 1.27$\pm$0.03,1.23$\pm$0.08 & 1,4 \\
WD\,0401+250 & 11 & 81.7 $\pm$ 0.3 & 10.16 & -0.37& 98.8 & 49.9 & 0.63$\pm$0.03 & 5 \\
WD\,0407+179 & 1  & 62.5$\pm$ 2.3 & --- &  --- & --- & --- & 0.49$\pm$0.03 & 5 \\
WD\,0416+334 & 6 & -44.4 $\pm$ 0.7 & 8.79 & -0.93& 90.7 & 45.2 &  \\
WD\,0416+701 & 18 & 21.3 $\pm$ 0.2 & 92.48 &-11.67& 98.7 & 62.6 &  \\
WD\,0437+152 & 8 & 21.1 $\pm$ 0.5 & 6.32 & -0.30& 95.5 & 48.3 & 0.38$\pm$0.03 & 5 \\
WD\,0446$-$789 & 7 & 40.2 $\pm$ 0.4 & 3.24 & -0.11& 91.5 & 42.0 & 0.51$\pm$0.02 7 7 \\
WD\,0453+418 & 15 & 59.7 $\pm$ 0.2 & 29.83 & -2.09& 99.0 & 69.3 & 0.43$\pm$0.03 & 1 \\
WD\,0507+045.1 & 6 & 37.8 $\pm$ 0.8 & 6.95 & -0.65& 90.2 & 32.3 & 0.61$\pm$0.03 & 5 \\
WD\,0507+045.2 & 6 & 48.2 $\pm$ 1.5 & 4.24 & -0.29& 85.4 & 15.2 & 0.71$\pm$0.03 & 5 \\
WD\,0509$-$007 & 5 & 22.1 $\pm$ 1.5 & 3.16 & -0.27& 88.5 & 36.1 & 0.382$\pm$0.005 & 3 \\
WD\,0516+365 & 2 & 54.8 $\pm$ 4.6 & 0.43 & -0.29& 26.0 & 1.9 & 0.59$\pm$0.03 & 5 \\
WD\,0549+158 & 17 & 30.0 $\pm$ 0.6 & 18.00 & -0.49& 95.7 & 38.9 & 0.47$\pm$0.02, 0.51$\pm$0.01 & 4,3 \\
WD\,0658+624 & 6 & 13.6 $\pm$ 0.7 & 6.33 & -0.56& 90.7 & 37.1 & 0.54$\pm$0.03 & 5 \\
WD\,0752$-$146 & 4 & 28.6 $\pm$ 1.2 & 19.19 & -3.60& 86.1 & 34.7 &  \\
WD\,0752$-$146\,B & 4 & -146.8 $\pm$ 1.3 & 17.58 & -3.27& 85.3 & 33.4 &  \\
WD\,0808+595 & 7 & 15.3 $\pm$ 1.1 & 7.62 & -0.57& 88.9 & 29.6 & 0.37$\pm$0.03 & 5 \\
WD\,0824+288\,B & 2 & -36.8 $\pm$ 2.9 & 0.02 & -0.06& 19.3 & 3.6 &  \\
WD\,0839+231 & 8 & 0.3 $\pm$ 0.5 & 3.72 & -0.09& 93.0 & 40.7 & 0.48$\pm$0.03,0.48$\pm$0.01 & 1,3 \\
WD\,0906+296 & 8 & 93.5 $\pm$ 1.0 & 3.95 & -0.11& 84.5 & 23.9 & 0.52$\pm$0.03 & 5 \\
WD\,0913+442 & 6 & 58.6 $\pm$ 0.7 & 5.93 & -0.50& 91.2 & 33.1 & 0.76$\pm$0.04,0.70$\pm$0.03 & 2,5 \\
WD\,0945+245 & 5 & 62.7 $\pm$ 2.1 & 2.78 & -0.23& 82.9 & 27.1 &  \\
WD\,0950$-$572 & 2 & 46.2 $\pm$ 6.7 & 0.01 & -0.04& 27.8 & 3.3 & 0.42$\pm$0.03 & 5 \\
WD\,0954+247 & 5 & 59.9 $\pm$ 0.7 & 6.91 & -0.85& 85.4 & 37.3 &  \\
WD\,0954$-$710 & 7 & 18.6 $\pm$ 0.3 & 16.86 & -2.01& 95.5 & 53.8 & 0.47$\pm$0.03,0.45$\pm$0.04 & 5,7 \\
WD\,1026+023 & 8 & 18.2 $\pm$ 0.6 & 11.05 & -0.86& 90.7 & 37.8 & 0.53$\pm$0.03,0.54$\pm$0.03 & 3,5 \\
WD\,1029+537 & 5 & 35.4 $\pm$ 5.6 & 12.46 & -1.85& 61.1 & 1.0 & 0.58$\pm$0.02 & 3 \\
WD\,1031$-$114 & 8 & 41.1 $\pm$ 0.9 & 5.54 & -0.23& 97.7 & 48.2 & 0.52$\pm$0.01, 0.57$\pm$0.03 & 3,5 \\
WD\,1036+433 & 5 & -5.6 $\pm$ 0.4 & 2.19 & -0.15& 97.3 & 64.3 &  \\
WD\,1039+747 & 4 & 47.7 $\pm$ 3.8 & 2.88 & -0.39& 60.8 & 6.8 & 0.45$\pm$0.03 & 1 \\
WD\,1105$-$048 & 18 & 50.8 $\pm$ 0.1 & 36.10 & -2.35& 99.9 & 88.9 & 0.49$\pm$0.03,0.48$\pm$0.03,0.53$\pm$0.03 & 1,4,5 \\
WD\,1229$-$012 & 9 & 18.6 $\pm$ 1.0 & 26.10 & -3.00& 95.4 & 38.5 & 0.42 $\pm$0.03 & 5 \\
WD\,1232+479 & 10 & 6.0 $\pm$ 0.4 & 17.27 & -1.35& 91.1 & 40.1 & 0.53$\pm$0.03 & 1 \\
WD\,1257+032 & 17 & 23.8 $\pm$ 0.4 & 25.99 & -1.27& 97.6 & 45.6 & 0.46$\pm$0.03 & 1 \\
WD\,1310+583 & 15 & 4.5 $\pm$ 0.3 & 8.97 & -0.08& 97.1 & 38.6 &  \\
WD\,1327$-$083 & 19 & 45.1 $\pm$ 0.1 & 82.77 & -9.55& 100.0 & 97.1 & 0.52$\pm$0.03,0.50$\pm$0.02 & 5,7 \\
WD\,1353+409 & 13 & -2.6 $\pm$ 0.5 & 16.29 & -0.75& 97.6 & 47.1 & 0.40$\pm$0.03 & 1 \\
WD\,1407$-$475 & 17 & 38.5 $\pm$ 0.2 &292.70 &$<-45$ & 99.2 &  64.7 &  0.50$\pm$0.02 & 7 \\
WD\,1422+095 & 6 & 1.6 $\pm$ 0.7 & 14.85 & -1.96& 92.1 & 41.3 & 0.51$\pm$0.04 & 7 \\
EUVE\,1439+750 & 4 & -140.9 $\pm$ 10.5 & 12.20 & -2.17& 27.7 & 2.1 & 0.96$\pm$0.05,0.99$\pm$0.05 & 6,6 \\
WD\,1507+220 & 10 & -50.7 $\pm$ 0.5 & 6.31 & -0.15& 93.3 & 42.2 & 0.50$\pm$0.03 & 1 \\
WD\,1507$-$105 & 2 & -14.0 $\pm$ 5.6 & 0.00 & -0.02& 26.9 & 6.1 &  \\
\end{tabular}
\end{table*}
\addtocounter{table}{-1}
\begin{table*}
\caption{continued.}
\begin{tabular}{@{}lrrrrrrlr}
Name & \multicolumn{1}{l}{N}&\multicolumn{1}{l}{Mean} & 
\multicolumn{1}{l}{$\chi^2$} & $\log_{10}(p)$ & \multicolumn{1}{l}{$e(A)$}& 
\multicolumn{1}{l}{$e$(B)} & Mass &\multicolumn{1}{l}{Ref.} \\
 & & \multicolumn{1}{l}{(km\,s$^{-1}$)}  & & & \multicolumn{1}{l}{(\%)} &
\multicolumn{1}{l}{(\%)} & (\Msolar) \\
WD\,1614+136 & 15 & 5.2 $\pm$ 0.5 & 25.67 & -1.54& 98.1 & 56.5 & 0.33$\pm$0.03 & 1 \\
WD\,1615$-$157 & 2 & 12.8 $\pm$ 6.7 & 0.13 & -0.14& 26.3 & 1.7 & 0.62$\pm$0.02,0.66$\pm$0.02 & 3,7 \\
WD\,1620$-$391 & 11 & 47.5 $\pm$ 0.1 & 32.44 & -3.47& 99.5 & 74.0 & 0.62$\pm$0.01,0.66$\pm$0.02 & 3,7 \\
WD\,1637+335 & 5 & 27.5 $\pm$ 1.0 & 8.80 & -1.18& 82.7 & 34.0 &  \\
WD\,1647+591 & 13 & 41.6 $\pm$ 1.1 & 7.71 & -0.09& 89.8 & 41.0 &  \\
HS\,1653+7753 & 5 & -1.2 $\pm$ 2.5 & 0.96 & -0.04& 71.6 & 16.9 & 0.32$\pm$0.02 & 4 \\
WD\,1655+215 & 5 & 40.0 $\pm$ 1.3 & 6.48 & -0.78& 87.2 & 34.3 &  \\
WD\,1911+135 & 10 & 20.7 $\pm$ 0.4 & 12.23 & -0.70& 94.0 & 45.6 & 0.49$\pm$0.03,0.50$\pm$0.03 & 1,2 \\
WD\,1943+163 & 14 & 36.3 $\pm$ 0.4 & 5.84 & -0.02& 97.4 & 50.5 & 0.49$\pm$0.03 & 1 \\
WD\,2058+506 & 15 & 8.2 $\pm$ 0.5 & 17.65 & -0.65& 92.2 & 44.5 &  \\
WD\,2111+261 & 12 & -2.4 $\pm$ 0.3 & 15.82 & -0.83& 92.5 & 49.4 &  \\
WD\,2117+539 & 11 & 3.3 $\pm$ 0.3 & 12.23 & -0.57& 95.8 & 53.2 & 0.50$\pm$0.03 & 1 \\
WD\,2136+828 & 10 & -35.6 $\pm$ 0.4 & 5.21 & -0.09& 94.4 & 46.8 & 0.50$\pm$0.03 & 1 \\
WD\,2151$-$015 & 11 & 41.0 $\pm$ 0.7 & 14.97 & -0.88& 95.6 & 44.7 &  \\
WD\,2151$-$015B & 4 & 6.8 $\pm$ 3.9 & 5.09 & -0.78& 70.0 & 18.8 &  \\
WD\,2226+061 & 10 & 40.7 $\pm$ 0.6 & 6.05 & -0.13& 92.2 & 40.8 & 0.43$\pm$0.03 & 1 \\
WD\,2341+322 & 5 & 7.5 $\pm$ 1.7 & 6.44 & -0.77& 84.5 & 22.7 & 0.57$\pm$0.03 & 5 \\
\end{tabular}
\end{table*}

\section{Notes on individual objects} \begin{description}
\item[WD\,0101+048:]{This object is variable according to our criterion. The
H$\alpha$ line is narrower than usual so we only use data within
2000\,km\,s$^{-1}$ of H$\alpha$  to measure the radial velocities. The
periodogram of these velocities is complex with peaks near 0.16cycles/d and
0.85cycles/d. We used a circular orbit fit by least squares to the measured
radial velocities to fix the position of the absorption core in each spectrum
in a least-squares fit to all the spectra to re-determine the model line
profile. We then re-measured the radial velocities with this improved model
line profile. These are the velocities given in Table~\ref{RVTable}. The
periodogram of these data shows  many  peaks of similar significance. We used
the seven most significant peaks  to give an initial value of the period in a
least-squares fit of a circular orbit to the data. The results are given in
Table~\ref{0101Table}. Periods near 6.4d and 1.2d are equally likely and there
are several periods near these values which would give a satisfactory fit to
our data. The real period of this binary should be easy to identify with a few
more spectra. The value of chi-squared is unusually low for all the circular
orbit fits in Table~\ref{0101Table}. There are 14 data points and 4 free
parameters in the fitting process, so we might expect a typical value of
chi-squared around 10, but a value of chi-squared as low as 5.27 occurs by
chance for about 1/8 trials, so such a low value of chi-squared is to be
expected occasionally given the number of binaries we have studied. }
\begin{figure} \leavevmode\centering{
\psfig{file=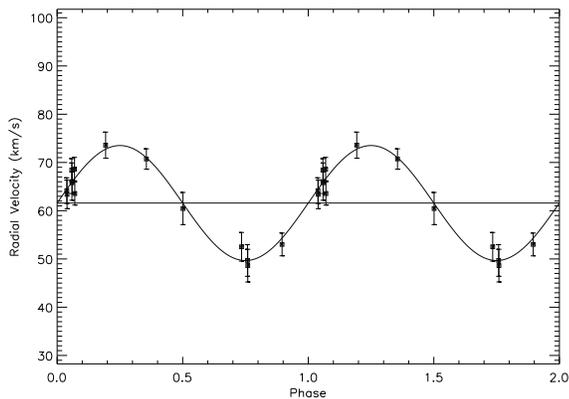,width=0.45\textwidth} } \caption{\label{0101Fig}
Measured radial velocities of WD\,0101+048 and our best circular orbit fit
(P=6.539d).} \end{figure}
\item[WD\,0341+021:]{ This star is clearly variable according to our
criterion. The variability is due to an offset between two data sets 
(see Table~\ref{OffsetTable}). Although there is only one spectrum in the INT
data set, the offset is clearly seen in the data and far exceeds the typical
offset between data sets. We suspect this is a long period binary.}
\item[WD\,0346$-$011:]{The H$\alpha$ line of this star is weak and broad so we
only used two Gaussians in the model line profile.}
\item[WD\,0416+701:]{This star is certainly variable according to our
criterion. We used the same procedure as for WD\,0101+048 to re-calculate the
model profile. There is a clear peak in the periodogram near 0.32d with no
other significant peaks. A circular orbit fit to the measured radial
velocities is given in Table~\ref{0416Table}. The $\chi^2$ value for this fit
is rather high so we present this as a tentative identification of the orbital
period. The measured radial velocities and circular orbit fit are shown in
Fig.~\ref{0416Fig}}
\begin{figure} \leavevmode\centering{
\psfig{file=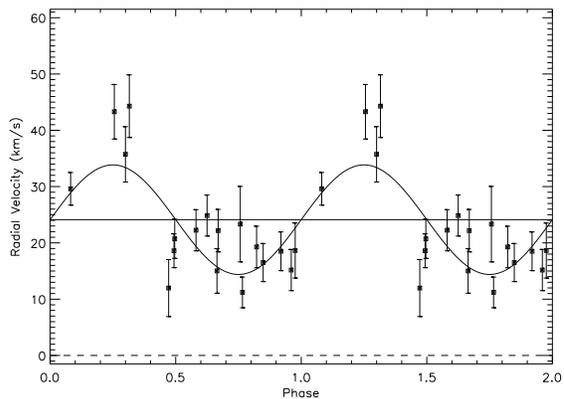,width=0.45\textwidth} } \caption{\label{0416Fig}
Measured radial velocities of WD\,0416+701 and a circular orbit fit.}
\end{figure}
\item[WD\,0752$-$146:]{Schultz et~al. (1992) found an emission line
superimposed on the usual absorption line which shows variable radial velocity
and indicates the presence of a companion. We measured the radial velocity of
both the absorption and emission lines and found both to be slightly variable,
though neither satisfies our strict criterion for binarity. The measurements of
the emission line are listed in Tables \ref{ResultsTable} and \ref{RVTable}
under WD\,0752$-$146\,B. We were unable to identify a definite period from our
data combined with the data of Schultz et~al.}
\item[WD\,0945+245:]{This star, also known as LB\,11146, was studied by Glenn
et~al. (1994) who found that the spectrum is a composite of a magnetic and a
non-magnetic white dwarf. Their radial velocity measurements showed no
variability  over a baseline of 16 days. We find no evidence for variability
from our own data nor from the  combination of both sets of radial velocity
measurements.}
\item[WD\,0824+288:]{This is a rare DA+dC star (Finley et~al. 1997) also known
as PG\,0824+289. We were unable to measure the radial velocity of the white
dwarf from our 2 H$\alpha$ spectra, but the results of measuring the radial
velocity of the dC component measured from the H$\alpha$ emission line are
given in Tables \ref{ResultsTable} and \ref{RVTable} under WD\,0824+288\,B. } 
\item[WD\,1029+537:]{This hot white dwarf has a broad, shallow H$\alpha$ line
so we only used two Gaussians to form the model line profile.}
\item[WD\,1036+433:]{The core of the H$\alpha$ line in star this reversed (in
emission).}
\item[WD\,1105$-$048:]{Although this star is variable according to our
criterion, there are no obvious periods in the data and circular orbit fits
to potential periods are not convincing.  The uncertainties for several of the
radial velocities given in Table~\ref{RVTable} are very small. These
uncertainties take no account of systematic errors in the data. If we assume
there is an additional uncertainty of only 0.5\,km\,s$^{-1}$ in the data, we
find $\log_{10}(p) = -3.3$. We believe that the data for this star simply
reflect the fact that systematic errors in the data limit the accuracy of our
radial velocity measurements to $\approx 0.5$\,km\,s$^{-1}$.}
\item[WD\,1310+583:]{We only used data within 3500\,km\,s$^{-1}$ of H$\alpha$
for  the fitting process because
four of our spectra only extend 3500\,km\,s$^{-1}$ to the
red of H$\alpha$.}
\item[WD\,1327$-$083:]{Although this star is variable according to our
criterion, there are no obvious periods in the data and circular orbit fits
to potential periods are not convincing. This would appear to be a similar
case to WD\,1105$-$048, i.e, the small uncertainties on some radial velocity
measurements are over-optimistic.}
\item[WD\,1407$-$475]{ This star is clearly variable according to our
criterion. The variability is due to the large offset between two data sets 
(see Table~\ref{OffsetTable}). A periodogram shows significant peaks near 1, 2 
and 3 cycles/d. We used the same technique applied to WD\,0101+048 to measure
the circular orbit fits to these three orbital periods given in
Table~\ref{1407Table}. The fit for a period near one day is shown in
Fig.~\ref{1407Fig}. The circular orbit fits are good but further data is
required to confirm this star is a binary and to then identify the correct
orbital period.}
\begin{figure} 
\leavevmode\centering{\psfig{file=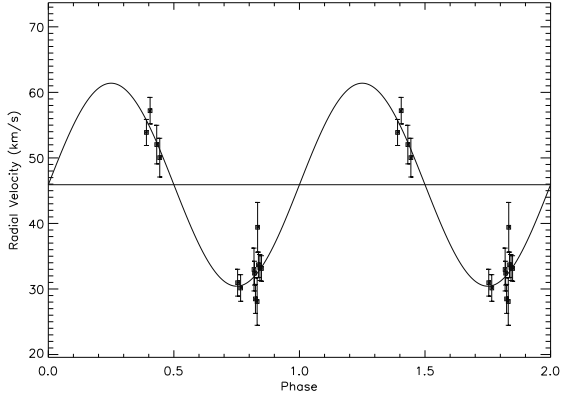,width=0.45\textwidth} } 
\caption{\label{1407Fig} Measured radial velocities of WD\,1407$-$475 and our
circular orbit fit for a period near one day.} 
\end{figure}
\item[WD\,1615$-$157:]{This star incorrectly labelled as 1615$-$154 by Bragaglia
et~al. (1995) and Saffer, Livio \& Yungelson (1998).}
\item[WD\,2151$-$015:]{We found this star showed
emission at H$\alpha$ due to a companion which is variable in strength (Maxted
et~al. 1999). We used an additional Gaussian component to model the emission
line, though it is not always visible in the spectra. Radial velocities for
the 4 spectra where the line could be measured are listed in Tables
\ref{ResultsTable} and \ref{RVTable} under WD\,2151$-$015\,B.} 
\end{description}

\begin{table}
\caption{\label{0101Table} Circular orbit fits to measured radial velocities of
WD\,0101+048. The uncertainty on the final digit of the period is given in
parentheses.}
\begin{tabular}{@{}rrrrr}
\multicolumn{1}{l}{Period} & \multicolumn{1}{l}{HJD(T$_0$)}\ &
\multicolumn{1}{l}{$\gamma$} & \multicolumn{1}{l}{K} &
 \multicolumn{1}{l}{$\chi^2$} \\
\multicolumn{1}{l}{(d)} &\multicolumn{1}{l}{$-$2451000} &
\multicolumn{1}{l}{(km\,s$^{-1}$)} & \multicolumn{1}{l}{(km\,s$^{-1}$)} \\
 6.539(4) & $ 3.9 \pm 0.1 $& $61.6 \pm 1.1$ & $11.9 \pm 1.3$ & 5.27 \\
 6.272(3) & $ 0.3 \pm 0.1 $& $62.2 \pm 0.7$ & $12.1 \pm 1.3$ & 5.56 \\
 5.807(3) & $ 5.7 \pm 0.1 $& $61.0 \pm 0.8$ & $10.9 \pm 1.1$ & 7.07 \\
 5.304(3) & $ 5.9 \pm 0.1 $& $62.0 \pm 1.0$ & $11.8 \pm 1.3$ & 6.70 \\
1.2093(2) & $6.10 \pm 0.03$& $62.3 \pm 0.7$ & $10.9 \pm 1.1$ & 6.49 \\
1.1768(1) & $6.65 \pm 0.03$& $61.0 \pm 0.8$ & $10.7 \pm 1.1$ & 5.88 \\
1.1461(1) & $7.19 \pm 0.03$& $59.1 \pm 0.9$ & $11.7 \pm 1.3$ & 7.02 \\
\end{tabular}
\end{table}

\begin{table}
\caption{\label{0416Table} A circular orbit fit to the measured radial
velocities of WD\,0416+701. The uncertainty on the final digit of the period
is given in parentheses.}
\begin{tabular}{@{}rrrrr}
\multicolumn{1}{l}{Period} & \multicolumn{1}{l}{HJD(T$_0$)}\ &
\multicolumn{1}{l}{$\gamma$} & \multicolumn{1}{l}{K} &
 \multicolumn{1}{l}{$\chi^2$} \\
\multicolumn{1}{l}{(d)} &\multicolumn{1}{l}{$-$2451000} &
\multicolumn{1}{l}{(km\,s$^{-1}$)} & \multicolumn{1}{l}{(km\,s$^{-1}$)} \\
0.31854(2) &  0.08$\pm$ 0.01& 24.9 $\pm$ 1.0&11.5 $\pm$ 1.6 & 30.0 \\
\end{tabular}
\end{table}

\begin{table}
\caption{\label{1407Table} Circular orbit fits to measured radial velocities of
WD\,1407$-$475.  The uncertainty on the final digit of the period
is given in parentheses.}
\begin{tabular}{@{}rrrrr}
\multicolumn{1}{l}{Period} & \multicolumn{1}{l}{HJD(T$_0$)}\ &
\multicolumn{1}{l}{$\gamma$} & \multicolumn{1}{l}{K} &
 \multicolumn{1}{l}{$\chi^2$} \\
\multicolumn{1}{l}{(d)} &\multicolumn{1}{l}{$-$2450000} &
\multicolumn{1}{l}{(km\,s$^{-1}$)} & \multicolumn{1}{l}{(km\,s$^{-1}$)} \\
 0.9985(3)  & $595.7  \pm 0.1  $& $46.2 \pm 5.4$ & $15.5 \pm 5.2$ & 10.7 \\
 0.50032(5) & $595.10 \pm 0.04 $& $42.9 \pm 1.4$ & $12.8 \pm 1.4$ & 10.7 \\
 0.33320(1) & $595.76 \pm 0.02 $& $41.8 \pm 0.8$ & $13.3 \pm 0.9$ & 11.9 \\
\end{tabular}
\end{table}

\section{Discussion}

\subsection{Evidence for a population of single, low mass white dwarfs.}
 We have used the results in Table~\ref{ResultsTable} to investigate whether
there is any evidence for a population of single, LMWDs. 
For the purposes of
this discussion we define an LMWD to be a white dwarf for which  more than
half the mass estimates, $M\pm \sigma_M$, satisfy the condition $(M_{\rm
lim} - M) > 2\sigma_M$, i.e., they are at least two standard deviations
below some mass limit $M_{\rm lim}$. The value of $M_{\rm lim}$ is a matter of
some debate, so we consider three cases, $M_{\rm lim}=0.45\Msolar$ ,$M_{\rm
lim}=0.50\Msolar$ and $M_{\rm lim}=0.55\Msolar$. Of the 20 white dwarfs which
satisfy the condition $M_{\rm lim}=0.55\Msolar$, 19 show no evidence for a
binary companion. The nature of the companion to WD\,0341+021, if it is a
binary,  is not known. The results in Table~\ref{ResultsTable} cannot be taken
at face value because the obvious binaries  have already been excluded. In
order to account for these binaries we have reviewed our records to identify
objects excluded from Table~\ref{ResultsTable} which were observed because of
their low mass and subsequently discovered to be binaries. There are 16 such
stars, 3 of which have main-sequence companions and 13 of which are known or
strongly suspected to have white dwarf companions. 

 It must be emphasized that we did not set out from the start to observe white
dwarfs is such a way as to determine whether there is any evidence for a
population of single LMWDs. The various stars were observed for different
reasons, sometimes with a different motivation  for the same star at different
times. Nevertheless, these stars were, in general, observed because of their
low mass and we continued to observe them if possible until we had either
established an orbital period or had established that they were likely to be
single. Therefore, our sample of LMWDS is fairly homogeneous and while it
is not ideal it is, by far, the best available. We have simplified our
analysis by assuming that there are only two populations of binary LMWDs,
those with a companion of equal mass (population $A$) and those with
companions of mass 0.08\Msolar (population $B$) and that our detection
efficiencies for these binaries are as given in Table~\ref{ResultsTable}. The
question we address here is whether our data show evidence for a population of
single LMWDs (population $C$). We then have two models. The first model is
that all LMWDs belong to either population $A$ or population $B$. We denote
this model M2 because it contains only two populations. The second model is
that there are three populations of LMWDs, $A$, $B$ and $C$, so we denote this
model M3. Using Bayes' theorem we find:
\[ \frac{\rm P(M3|D)}{\rm P(M2|D)} = \frac{\rm P(M3)}{\rm P(M2)} \frac{\rm
P(D|M3)}{\rm P(D|M2)},  \] 
where the usual notation applies, e.g., \({\rm P(M2|D)}\) is the probability
of model M2 given our data, D. Our data consist of $N_A$ LMWDs with white
dwarf companions we identify as belonging to population $A$, $N_B$ LMWDS with
main-sequence companions we identify as belonging to population $B$ and $N_0$
that are not detected as binaries. For these ``non-detections'', we have
detection efficiencies $e_i(A)$ and $e_i(B)$, $i=1,2,\dots,N_0$, for binaries
belonging to population $A$ and $B$, respectively. If some fraction $f_A$ of
binaries belong to population $A$ and some fraction $f_C$ of all LMWDs belong
to population $C$, then
\[\frac{\rm P(D|M3)}{\rm P(D|M2)} =
\frac{((1-f_C)f_A)^{N_A}((1-f_C)(1-f_A))^{N_B}\prod q_i} {
f_A^{N_A}(1-f_A)^{N_B}\prod p_i} \]
where
\[ p_i = f_A(1-e_i(A)) + (1-f_A)(1-e_i(B)) \]
and
\[ q_i = f_A(1-e_i(A)) + (1-f_A)(1-e_i(B))+f_C \]
 This ratio of probabilities has the considerable merit that we do not need to
make any assumptions concerning the detection efficiencies for those LMWDs
identified as belonging to population $A$ or $B$. As we have no prior
assumptions concerning the values of $f_A$ or  $f_C$, we simply integrate the
function numerically over a uniform grid of all possible values. In addition,
we can identify the most likely values of $f_A$ and $f_C$ given our data.

 We have calculated the value of $\frac{P(D|M3)}{P(D|M2)}$ for the three cases
shown in Table~\ref{FracTable} corresponding to three different assumptions
concerning the nature of the companion to WD\,0341+021. We include the case of
an undetected companion despite having noted this star as a binary to allow
for this detection being a ``false-alarm'', though we consider this to be
unlikely. Also given in Table~\ref{FracTable} are the most likely values of
$f_A$ and $f_C$.

\begin{table}
\caption{\label{FracTable} The ratio of probabilities
$\frac{P(D|M3)}{P(D|M2)}$ for three different assumptions concerning
WD\,0341+021 and the most likely values of $f_A$ and $f_C$ and for three
different uper limits to the mass, $M_{\rm lim}$. The number of
stars from Table~\ref{ResultsTable} whose measured masses are two standard
deviations below $M_{\rm lim}$, $N_{\rm low}$, is also given.}
\begin{tabular}{lrrrrr}
\multicolumn{1}{l}{Companion to} &\multicolumn{1}{l}{\underline{$P(D|M3)$}} &
\multicolumn{1}{l}{Model M2} & \multicolumn{2}{l}{Model M3} \\
\multicolumn{1}{l}{~WD\,0341+021}
&\multicolumn{1}{l}{$P(D|M2)$} &
\multicolumn{1}{c}{$f_A$} & \multicolumn{1}{c}{$f_A$}&
\multicolumn{1}{c}{$f_C$} \\
\multicolumn{4}{l}{$M_{\rm lim}$=0.55\Msolar, $N_{\rm low} = 20$} \\
~Undetected & 27   & 0.42 & 0.67 & 0.39 \\
~White dwarf & 16   & 0.45 & 0.68 & 0.36 \\
~Main sequence & 7    & 0.42 & 0.60 & 0.33 \\
\multicolumn{4}{l}{$M_{\rm lim}$=0.50\Msolar, $N_{\rm low} = 14$} \\
~Undetected & 48   & 0.49 & 0.68 & 0.30 \\
~White dwarf & 30   & 0.52 & 0.70 & 0.27 \\
~Main sequence & 17   & 0.48 & 0.62 & 0.23 \\
\multicolumn{4}{l}{$M_{\rm lim}$=0.45\Msolar, $N_{\rm low} = 8 $} \\
~Undetected & 43   & 0.60 & 0.67 & 0.12 \\
~White dwarf & 30   & 0.64 & 0.69 & 0.08 \\
~Main sequence & 23   & 0.59 & 0.61 & 0.03 \\
\end{tabular}
\end{table}

The sensitivity of our result to the assumed properties of just one star in a
sample of 37 demonstrates that the results must be treated with some caution. 
However, they do seem to favour the existence of a population of single LMWDs.
This result should not be taken as conclusive for several reasons. Firstly, we
have assumed that some of the binaries from our other studies which show no
sign of a companion star have white dwarf companions. If this is not the case,
then the value $\frac{P(D|M3)}{P(D|M2)}$ may be much lower. Secondly, there
have been many simplifying assumptions made concerning the nature of the
companions to these star. Thirdly, we have used the theoretical period
distribution for these binaries despite the problems with these theories. In
summary, we can say the the data favour the existence of a population of
single LMWDs, but this result is not conclusive.

We have assumed that the companions to the LMWDs in our sample are either
white dwarfs or main-sequence stars ($M\ge 0.08\Msolar$). The question of
whether companions of such low mass or lower (i.e., sub-stellar companions)
can survive a common envelope phase is a difficult one to answer (Siess \&
Livio 1999). It would certainly be useful to continue observations of the
LMWDs presented here to push down the limits on the mass of any possible
companion.

\subsection{Comparison with the results of Saffer, Livio \& Yungelson (1998).}
Several of the stars in this paper are candidate radial velocity variables
from Saffer , Livio \& Yungelson (1998). There are four ``weight 1''
candidates (WD\,1232+479, WD\,1310+583, WD\,1647+591, WD\,2117+539) and four
``weight 2'' candidates (WD\,0401+250, WD\,0549+158, WD\,0839+231,
WD\,1229$-$012). Only one of these shows any hint of variability from our own
data, which is quite extensive for all these stars. This is, perhaps, not
surprising given that Maxted \& Marsh (1999) found that the mean number of
false detections of binaries  expected in their survey based on the quoted
uncertainty in the radial velocity measurements and the detection criterion is
17.7. This estimate  is clearly too high given the number of binary candidates
identified by the survey which were known to be binaries beforehand or which
have been confirmed subsequently. This suggests that the typical uncertainty
 quoted for these radial velocity measurements is to low. It also shows the
problems that can arise when trying to draw quantitative conclusions from a
survey for binary stars based on rather subjective detection criteria. 

\section{Conclusion}
 We have presented 594 radial velocity measurements for 71 white dwarfs. We
find that WD\,0101+048 is certainly a binary, but are unable to determine
whether the orbital period is near 6.4d or 1.2d. Similarly,  WD\,1407$-$475 is
also a binary but we are unable to determine whether its orbital period is
near 1d, 1/2d or 1/3d from our data. WD\,0416+701 is likely to be binary and
our data favours an orbital period of 0.32d, but further observations are
required to show this convincingly. We also identify WD\,0341+021 as another
likely binary but are unable to establish the orbital period fom our data.
There is some evidence in our data for a population of single, low mass white
dwarfs, but this result is dependent on several assumptions.

\begin{table}
\caption{\label{RVTable} Measured heliocentric radial velocities.}
\begin{tabular}{@{}lrr}
Name & \multicolumn{1}{l}{HJD} & \multicolumn{1}{l}{Radial velocity}  \\
& \multicolumn{1}{l}{-2400000} & (km\,s$^{-1}$) \\
EUVE1439+750 \\
& 51241.6338&  -144.2$\pm$  27.6\\
& 51241.6998&   -79.4$\pm$  25.6\\
& 51242.7223&  -170.4$\pm$  35.4\\
& 51242.7370&  -227.4$\pm$  36.2\\
HS1653+7753 \\
& 51067.3966&    -3.2$\pm$  11.1\\
& 51067.4588&    -0.8$\pm$  11.7\\
& 51067.4730&    -5.0$\pm$  11.0\\
& 51243.6822&    -4.9$\pm$   7.4\\
& 51243.7039&     4.0$\pm$   6.9\\
WD\,0011+000 \\
& 51071.5057&    18.4$\pm$   5.7\\
& 51071.5198&    28.5$\pm$   3.6\\
& 51071.6775&    29.5$\pm$   4.0\\
WD\,0101+048 \\
& 50760.3534&    49.7$\pm$   3.3\\
& 50760.3606&    48.6$\pm$   3.4\\
& 50762.3176&    68.4$\pm$   2.4\\
& 50762.3271&    65.8$\pm$   2.2\\
& 50762.3885&    68.6$\pm$   2.4\\
& 50762.3957&    63.5$\pm$   2.4\\
& 51067.5292&    52.5$\pm$   3.0\\
& 51068.5837&    53.0$\pm$   2.4\\
& 51070.5287&    73.6$\pm$   2.7\\
& 51071.5898&    70.7$\pm$   2.1\\
& 51072.5427&    60.4$\pm$   3.3\\
& 50677.1806&    64.2$\pm$   2.7\\
& 50677.1879&    63.4$\pm$   3.0\\
& 50677.3150&    66.0$\pm$   3.9\\
WD\,0126+101 \\
& 50777.4800&     7.3$\pm$   1.8\\
& 50777.4839&     6.3$\pm$   1.9\\
& 50777.5666&     4.8$\pm$   2.2\\
& 50777.5704&     3.6$\pm$   2.3\\
& 50778.5530&     8.9$\pm$   1.4\\
& 50778.5594&     8.4$\pm$   1.9\\
WD\,0132+254 \\
& 50854.3391&    34.1$\pm$   5.8\\
& 50854.3603&    25.8$\pm$   5.4\\
& 50775.3529&    36.3$\pm$   2.0\\
& 50775.3696&    36.4$\pm$   2.1\\
& 50775.5094&    38.2$\pm$   1.9\\
& 50775.5260&    37.2$\pm$   2.0\\
& 50777.3592&    34.6$\pm$   2.3\\
& 50777.3737&    36.1$\pm$   2.2\\
WD\,0142+312 \\
& 51067.5768&    38.8$\pm$   7.3\\
& 51067.5909&    52.2$\pm$   5.3\\
& 51067.6452&    35.7$\pm$  14.9\\
& 51067.6593&    30.8$\pm$   9.9\\
& 51067.7305&    39.0$\pm$   5.9\\
& 51068.5971&    30.5$\pm$   4.8\\
& 51068.6919&    32.2$\pm$   5.1\\
& 51069.7306&    38.0$\pm$   4.9\\
WD\,0143+216 \\
& 51071.5352&    27.6$\pm$   3.9\\
& 51071.5493&    13.7$\pm$   3.7\\
& 51071.6594&    22.7$\pm$   3.6\\
& 51072.5870&    19.8$\pm$   5.7\\
& 51072.7436&    18.0$\pm$   7.1\\
\\
\\
\\
\end{tabular}
\end{table}
\begin{table}
\contcaption{}
\begin{tabular}{@{}lrr}
Name & \multicolumn{1}{l}{HJD} & \multicolumn{1}{l}{Radial velocity}  \\
& \multicolumn{1}{l}{-2400000} & (km\,s$^{-1}$) \\
WD\,0147+674 \\
& 49742.3796&    30.1$\pm$   4.7\\
& 49742.3867&    38.5$\pm$   4.7\\
& 49742.3938&    22.3$\pm$   4.7\\
& 49739.3462&    25.9$\pm$   5.2\\
& 49739.3533&    26.1$\pm$   6.1\\
& 49739.3607&    40.0$\pm$   6.1\\
WD\,0148+467 \\
& 51067.7424&    14.1$\pm$   2.4\\
& 51067.7495&    16.4$\pm$   2.5\\
WD\,0151+017 \\
& 50778.5345&    64.0$\pm$   2.2\\
& 50778.5442&    62.8$\pm$   2.2\\
& 50778.6289&    61.7$\pm$   2.6\\
& 50778.6385&    64.3$\pm$   2.7\\
WD\,0213+396 \\
& 51070.5847&    56.6$\pm$  13.6\\
& 51070.6694&    22.8$\pm$   3.5\\
& 51070.6835&    27.2$\pm$   3.4\\
& 51070.7341&    19.9$\pm$   3.1\\
& 51072.5191&    54.3$\pm$   8.3\\
& 51072.7257&    28.9$\pm$   3.1\\
WD\,0316+345 \\
& 49742.4274&   -43.6$\pm$   1.8\\
& 49742.4333&   -44.0$\pm$   1.8\\
& 49742.4392&   -44.2$\pm$   1.9\\
& 49739.3721&   -39.1$\pm$   2.0\\
& 49739.3792&   -45.2$\pm$   2.0\\
& 49739.3863&   -43.1$\pm$   2.1\\
& 49738.4740&   -41.0$\pm$   2.4\\
& 49738.4870&   -42.4$\pm$   2.7\\
& 51067.6248&   -43.6$\pm$   4.7\\
& 51068.6430&   -46.6$\pm$   2.6\\
& 51068.6571&   -48.2$\pm$   2.7\\
& 51068.7065&   -32.9$\pm$   2.6\\
WD\,0320-539 \\
& 50143.9069&    61.3$\pm$   3.7\\
& 50143.9254&    63.7$\pm$   4.2\\
& 50144.9013&    52.1$\pm$   4.7\\
& 50144.9197&    44.3$\pm$   4.8\\
& 50144.9387&    64.4$\pm$   4.8\\
& 50145.9028&    55.8$\pm$   4.8\\
& 50145.9212&    58.9$\pm$   5.2\\
WD\,0332+320 \\
& 50778.5763&    94.9$\pm$   4.5\\
& 50778.5841&   103.8$\pm$   4.4\\
& 50778.6816&   103.3$\pm$   5.8\\
& 50778.6890&   103.5$\pm$   5.9\\
WD\,0339+523 \\
& 49742.4503&     5.8$\pm$   2.6\\
& 49742.4601&     4.4$\pm$   2.6\\
& 49742.4695&    -1.2$\pm$   2.6\\
& 49739.4000&     0.6$\pm$   3.4\\
& 49739.4123&     6.5$\pm$   2.8\\
& 49739.4280&     2.9$\pm$   2.8\\
& 49739.4420&     3.0$\pm$   2.9\\
& 49738.5090&    -2.2$\pm$   6.3\\
& 49738.5237&     7.3$\pm$   5.7\\
\\
\\
\\
\\
\\
\\
\end{tabular}
\end{table}
\begin{table}
\contcaption{}
\begin{tabular}{@{}lrr}
Name & \multicolumn{1}{l}{HJD} & \multicolumn{1}{l}{Radial velocity}  \\
& \multicolumn{1}{l}{-2400000} & (km\,s$^{-1}$) \\
WD\,0341+021 \\
& 50855.3520&   -22.3$\pm$   6.8\\
& 50775.4488&   -54.2$\pm$   2.1\\
& 50775.4631&   -53.6$\pm$   2.3\\
& 50775.6270&   -50.4$\pm$   2.4\\
& 50775.6413&   -49.3$\pm$   2.1\\
& 50777.4505&   -58.0$\pm$   2.2\\
& 50777.4649&   -57.3$\pm$   2.3\\
WD\,0346-011 \\
& 51068.7409&    89.6$\pm$  35.2\\
& 51069.7458&    70.4$\pm$  43.9\\
& 51070.7131&    99.8$\pm$  55.2\\
& 51070.7202&   179.9$\pm$  54.6\\
& 51071.6169&    42.2$\pm$  47.1\\
& 51071.6240&   310.4$\pm$  46.0\\
& 51071.6871&   120.5$\pm$  43.6\\
& 51071.6942&   121.1$\pm$  43.0\\
& 51071.7517&   225.3$\pm$  48.7\\
& 51071.7589&   136.0$\pm$  51.5\\
WD\,0401+250 \\
& 50855.3979&    83.4$\pm$   3.1\\
& 50855.4052&    80.1$\pm$   3.3\\
& 50760.6725&    85.2$\pm$   3.5\\
& 50760.6844&    83.4$\pm$   3.4\\
& 50762.7069&    84.6$\pm$   2.3\\
& 50762.7147&    79.3$\pm$   3.0\\
& 50778.5921&    80.9$\pm$   3.0\\
& 50778.5948&    75.5$\pm$   2.9\\
& 50778.5977&    84.3$\pm$   2.9\\
& 50778.7358&    79.0$\pm$   3.2\\
& 50778.7397&    81.6$\pm$   3.3\\
WD\,0407+179 \\
& 51243.3924&    62.5$\pm$   2.3\\
WD\,0416+334 \\
& 50775.5542&   -43.8$\pm$   2.1\\
& 50775.5638&   -43.4$\pm$   2.3\\
& 50776.7101&   -43.4$\pm$   4.5\\
& 50776.7198&   -50.7$\pm$   4.6\\
& 50777.6938&   -50.9$\pm$   3.1\\
& 50777.7058&   -40.7$\pm$   2.5\\
WD\,0416+701 \\
& 51067.6951&    43.9$\pm$   4.6\\
& 51067.7092&    36.1$\pm$   4.7\\
& 51068.5631&    18.6$\pm$   4.6\\
& 51068.6714&    44.1$\pm$   5.3\\
& 51068.7214&    11.7$\pm$   4.8\\
& 51070.6974&    15.3$\pm$   3.8\\
& 51070.7477&    19.4$\pm$   3.5\\
& 51071.6010&    21.0$\pm$   3.3\\
& 51071.7360&    18.6$\pm$   3.3\\
& 51072.6416&    24.0$\pm$   6.4\\
& 51072.7064&    15.2$\pm$   3.5\\
& 51240.4325&    22.3$\pm$   3.5\\
& 51240.4466&    25.0$\pm$   3.4\\
& 51240.4606&    22.3$\pm$   3.6\\
& 51241.4751&    16.1$\pm$   3.2\\
& 51242.4061&    11.3$\pm$   2.6\\
& 51243.4642&    29.6$\pm$   2.8\\
& 51271.3573&    18.6$\pm$   2.8\\
\\
\\
\\
\\
\\
\end{tabular}
\end{table}
\begin{table}
\contcaption{}
\begin{tabular}{@{}lrr}
Name & \multicolumn{1}{l}{HJD} & \multicolumn{1}{l}{Radial velocity}  \\
& \multicolumn{1}{l}{-2400000} & (km\,s$^{-1}$) \\
WD\,0437+152 \\
& 50854.3885&    20.1$\pm$   5.0\\
& 50854.4097&    22.6$\pm$   3.3\\
& 50775.5751&    20.9$\pm$   2.5\\
& 50775.5917&    19.0$\pm$   2.7\\
& 50777.5226&    22.1$\pm$   2.4\\
& 50777.5368&    24.9$\pm$   2.5\\
& 50777.6274&    15.5$\pm$   3.2\\
& 50777.6417&    20.9$\pm$   3.6\\
WD\,0446-789 \\
& 50143.9414&    42.7$\pm$   2.6\\
& 50143.9470&    38.6$\pm$   2.4\\
& 50144.9550&    39.3$\pm$   2.7\\
& 50144.9623&    40.5$\pm$   3.0\\
& 50145.9372&    37.9$\pm$   2.6\\
& 50145.9446&    40.4$\pm$   2.4\\
& 50145.9519&    42.6$\pm$   2.6\\
WD\,0453+418 \\
& 49742.4820&    57.5$\pm$   1.7\\
& 49742.4879&    56.9$\pm$   1.6\\
& 49742.4939&    60.0$\pm$   1.6\\
& 49742.5007&    57.8$\pm$   1.6\\
& 49739.4570&    59.4$\pm$   1.6\\
& 49739.4641&    60.0$\pm$   1.8\\
& 49739.4712&    59.9$\pm$   1.7\\
& 49738.5463&    60.5$\pm$   3.4\\
& 49738.5572&    61.6$\pm$   2.7\\
& 51238.3489&    52.0$\pm$   4.1\\
& 51238.4277&    62.6$\pm$   4.9\\
& 51238.4936&    52.4$\pm$   8.9\\
& 51240.3560&    59.2$\pm$   2.0\\
& 51241.3672&    62.9$\pm$   2.0\\
& 51242.4281&    67.2$\pm$   1.9\\
WD\,0507+045.1 \\
& 50775.6142&    36.6$\pm$   2.9\\
& 50775.6181&    36.3$\pm$   3.1\\
& 50776.7292&    33.7$\pm$   4.8\\
& 50776.7366&    35.2$\pm$   5.5\\
& 50777.5554&    45.9$\pm$   3.5\\
& 50777.5593&    36.6$\pm$   3.3\\
WD\,0507+045.2 \\
& 50775.6142&    43.4$\pm$   4.9\\
& 50775.6181&    47.0$\pm$   5.2\\
& 50776.7292&    59.6$\pm$   9.2\\
& 50776.7366&    49.3$\pm$  10.9\\
& 50777.5554&    44.7$\pm$   6.1\\
& 50777.5593&    55.1$\pm$   5.9\\
WD\,0509-007 \\
& 51240.4144&    24.2$\pm$   3.7\\
& 51240.4959&    18.0$\pm$  10.8\\
& 51241.4268&    26.1$\pm$   3.8\\
& 51242.3366&    15.9$\pm$   5.1\\
& 51242.4480&    20.8$\pm$   3.1\\
WD\,0516+365 \\
& 50778.6086&    52.7$\pm$   4.6\\
& 50778.6160&    57.0$\pm$   4.6\\
\\
\\
\\
\\
\\
\\
\\
\\
\end{tabular}
\end{table}
\begin{table}
\contcaption{}
\begin{tabular}{@{}lrr}
Name & \multicolumn{1}{l}{HJD} & \multicolumn{1}{l}{Radial velocity}  \\
& \multicolumn{1}{l}{-2400000} & (km\,s$^{-1}$) \\
WD\,0549+158 \\
& 50852.3862&    29.7$\pm$   4.9\\
& 50852.3969&    28.3$\pm$   4.1\\
& 50852.4088&    35.4$\pm$   4.7\\
& 50852.5026&    19.7$\pm$   5.5\\
& 50852.5099&    38.8$\pm$   8.2\\
& 50852.5138&    35.3$\pm$   7.9\\
& 50854.4282&    22.9$\pm$   6.4\\
& 50854.4320&    41.9$\pm$   6.3\\
& 50854.4358&    19.2$\pm$   6.3\\
& 50855.3795&    44.0$\pm$  26.2\\
& 50855.3868&    26.5$\pm$   6.7\\
& 51238.4681&    24.4$\pm$  10.2\\
& 51238.4787&    41.8$\pm$  13.5\\
& 51240.3339&    34.4$\pm$   6.4\\
& 51240.3411&    37.4$\pm$   6.3\\
& 51241.3499&    30.3$\pm$   4.2\\
& 51243.4199&    27.9$\pm$   4.6\\
WD\,0658+624 \\
& 50775.6884&    11.9$\pm$   3.5\\
& 50775.7015&     9.9$\pm$   3.0\\
& 50776.7544&    22.9$\pm$   4.9\\
& 50776.7710&    16.3$\pm$   3.5\\
& 50777.7215&    14.4$\pm$   2.8\\
& 50777.7359&    12.1$\pm$   3.1\\
WD\,0752-146 \\
& 51243.3630&    30.7$\pm$   4.1\\
& 51243.3717&    26.2$\pm$   4.0\\
& 51243.4397&    21.6$\pm$   2.7\\
& 51267.8694&    40.4$\pm$   3.5\\
WD\,0752-146B \\
& 51243.3630&  -152.2$\pm$   4.3\\
& 51243.3717&  -158.6$\pm$   4.2\\
& 51243.4397&  -138.9$\pm$   2.8\\
& 51267.8694&  -147.6$\pm$   3.7\\
WD\,0808+595 \\
& 50852.5339&     3.5$\pm$   8.3\\
& 50852.5559&    32.7$\pm$   8.7\\
& 50775.7134&    14.7$\pm$   5.3\\
& 50775.7300&     9.8$\pm$   7.4\\
& 50777.5917&    18.2$\pm$   6.1\\
& 50777.6060&    19.8$\pm$   6.7\\
& 50777.7610&    12.6$\pm$   4.7\\
WD\,0824+288B \\
& 50777.7806&   -36.5$\pm$   3.0\\
& 50777.7846&   -37.1$\pm$   2.7\\
WD\,0839+231 \\
& 49742.5339&    -2.1$\pm$   3.5\\
& 49742.5398&    -1.7$\pm$   3.5\\
& 49742.5457&     1.1$\pm$   3.6\\
& 49739.5844&     2.9$\pm$   3.4\\
& 49739.5916&    -0.1$\pm$   3.6\\
& 49739.5987&    -0.8$\pm$   3.4\\
& 49738.6393&    -1.1$\pm$   3.0\\
& 49738.6501&     5.8$\pm$   3.9\\
WD\,0906+296 \\
& 50775.6573&    94.2$\pm$   3.3\\
& 50775.6670&    90.8$\pm$   3.9\\
& 50775.7501&   111.4$\pm$  11.1\\
& 50775.7622&    96.1$\pm$   8.3\\
& 50775.7744&    93.8$\pm$   4.0\\
& 50775.7864&    93.5$\pm$   7.9\\
& 50777.6576&    93.8$\pm$   5.6\\
& 50777.6673&    86.5$\pm$   8.1\\
\end{tabular}
\end{table}
\begin{table}
\contcaption{}
\begin{tabular}{@{}lrr}
Name & \multicolumn{1}{l}{HJD} & \multicolumn{1}{l}{Radial velocity}  \\
& \multicolumn{1}{l}{-2400000} & (km\,s$^{-1}$) \\
WD\,0913+442 \\
& 50760.7469&    57.2$\pm$   3.0\\
& 50760.7645&    57.4$\pm$   3.1\\
& 50761.6966&    62.8$\pm$   3.7\\
& 50761.7119&    53.3$\pm$   3.9\\
& 50762.7240&    63.2$\pm$   3.1\\
& 50762.7393&    57.2$\pm$   3.1\\
WD\,0945+245 \\
& 51240.4008&    67.7$\pm$   9.5\\
& 51241.4105&    70.9$\pm$   7.1\\
& 51242.3625&    58.1$\pm$   8.4\\
& 51242.6266&    56.0$\pm$   8.0\\
& 51243.4934&    59.1$\pm$   8.4\\
WD\,0950-572 \\
& 51267.8817&    46.8$\pm$   6.8\\
& 51267.8958&    45.7$\pm$   6.6\\
WD\,0954+247 \\
& 50761.7309&    56.2$\pm$   2.6\\
& 50762.6712&    56.8$\pm$   2.9\\
& 50762.6853&    59.3$\pm$   2.9\\
& 50762.7556&    64.5$\pm$   2.6\\
& 50762.7692&    62.3$\pm$   2.7\\
WD\,0954-710 \\
& 50143.0878&    24.6$\pm$   2.6\\
& 50143.0998&    23.4$\pm$   1.7\\
& 50144.1095&    18.3$\pm$   1.7\\
& 50145.0226&    17.7$\pm$   1.5\\
& 50145.0300&    17.0$\pm$   1.5\\
& 50146.1061&    16.8$\pm$   1.4\\
& 50146.1134&    17.9$\pm$   1.4\\
WD\,1026+023 \\
& 50776.7813&    26.4$\pm$   4.4\\
& 50776.7851&    28.5$\pm$   4.4\\
& 50776.7890&    15.8$\pm$   4.7\\
& 50777.7908&    16.9$\pm$   2.4\\
& 50777.7947&    14.9$\pm$   3.4\\
& 50778.6587&    16.7$\pm$   3.2\\
& 50778.6627&    16.3$\pm$   3.1\\
& 50778.7867&    18.9$\pm$   4.7\\
WD\,1029+537 \\
& 51004.3892&    13.2$\pm$  23.6\\
& 51241.4523&   -19.6$\pm$  23.3\\
& 51242.5058&    54.9$\pm$  19.9\\
& 51242.5274&    79.7$\pm$  20.1\\
& 51243.5192&    24.7$\pm$  24.5\\
WD\,1031-114 \\
& 51238.5873&    44.9$\pm$   6.4\\
& 51238.6013&    35.2$\pm$   4.8\\
& 51241.5576&    43.3$\pm$   2.8\\
& 51241.5682&    40.6$\pm$   2.7\\
& 51242.5494&    39.0$\pm$   2.3\\
& 51243.5883&    40.3$\pm$   2.7\\
& 51267.9177&    57.7$\pm$  15.1\\
& 51268.0079&    43.4$\pm$   2.4\\
WD\,1036+433 \\
& 49162.3636&    -7.1$\pm$   1.5\\
& 49162.3692&    -6.1$\pm$   1.4\\
& 49153.3770&    -4.9$\pm$   1.9\\
& 49153.3815&    -4.3$\pm$   1.8\\
& 49150.3969&    -4.8$\pm$   1.4\\
\\
\\
\\
\\
\end{tabular}
\end{table}
\begin{table}
\contcaption{}
\begin{tabular}{@{}lrr}
Name & \multicolumn{1}{l}{HJD} & \multicolumn{1}{l}{Radial velocity}  \\
& \multicolumn{1}{l}{-2400000} & (km\,s$^{-1}$) \\
WD\,1039+747 \\
& 49742.5553&    34.0$\pm$   9.9\\
& 49742.5659&    56.0$\pm$   9.6\\
& 49739.6153&    52.5$\pm$  12.6\\
& 49739.6259&    51.0$\pm$  12.6\\
WD\,1105-048 \\
& 51241.5141&    50.4$\pm$   1.6\\
& 51241.5247&    48.3$\pm$   1.3\\
& 51241.6078&    50.7$\pm$   1.6\\
& 51241.6184&    49.8$\pm$   1.5\\
& 51241.7391&    56.2$\pm$   4.2\\
& 51242.6453&    52.1$\pm$   1.6\\
& 50527.1917&    50.4$\pm$   1.3\\
& 50527.1991&    50.0$\pm$   1.5\\
& 50528.1782&    57.1$\pm$   2.8\\
& 50528.1925&    49.8$\pm$   1.5\\
& 50143.1897&    53.3$\pm$   1.3\\
& 50143.1947&    53.3$\pm$   0.3\\
& 50144.1248&    51.0$\pm$   1.6\\
& 50144.1309&    50.7$\pm$   1.2\\
& 50145.1577&    50.7$\pm$   0.3\\
& 50145.1627&    50.6$\pm$   0.4\\
& 50146.1267&    48.9$\pm$   0.8\\
& 50146.1318&    49.4$\pm$   0.3\\
WD\,1229-012 \\
& 51238.6209&    27.8$\pm$   6.5\\
& 51238.6349&    14.9$\pm$   9.4\\
& 51238.7430&   -27.9$\pm$  13.3\\
& 51238.7571&    31.5$\pm$  11.5\\
& 51240.6630&    23.7$\pm$   3.7\\
& 51240.7640&     5.5$\pm$   5.3\\
& 51243.6141&    15.4$\pm$   3.4\\
& 51268.0433&    21.6$\pm$   3.6\\
& 51268.1519&    25.6$\pm$   7.4\\
WD\,1232+479 \\
& 50852.7120&    -6.6$\pm$   5.0\\
& 50852.7333&     2.4$\pm$   3.2\\
& 50853.6397&     9.5$\pm$   2.8\\
& 50853.6488&     7.5$\pm$   4.2\\
& 50853.7267&     9.0$\pm$   3.9\\
& 50853.7339&     8.1$\pm$   3.9\\
& 50854.5369&     0.9$\pm$   2.8\\
& 50854.5501&     3.6$\pm$   3.1\\
& 50854.6129&     9.5$\pm$   3.1\\
& 50854.6248&     9.8$\pm$   2.8\\
WD\,1257+032 \\
& 49888.4357&    34.3$\pm$   5.2\\
& 49888.4578&    21.5$\pm$   5.3\\
& 49889.4251&    28.4$\pm$   4.1\\
& 49889.4471&    12.9$\pm$   4.4\\
& 49891.4325&    25.0$\pm$   5.1\\
& 49891.4551&    25.0$\pm$   4.7\\
& 49892.3956&    22.4$\pm$   6.1\\
& 49892.4290&     4.5$\pm$   6.6\\
& 49893.4011&    25.3$\pm$   7.7\\
& 49893.4230&    18.9$\pm$   7.3\\
& 49742.7052&    24.4$\pm$   3.4\\
& 49742.7157&    28.6$\pm$   3.2\\
& 49742.7279&    24.4$\pm$   3.2\\
& 51243.6395&    14.4$\pm$   8.6\\
& 51268.0664&    26.6$\pm$   9.4\\
& 51268.1739&    14.9$\pm$   8.6\\
& 51268.1880&    31.6$\pm$  10.1\\
\end{tabular}
\end{table}
\begin{table}
\contcaption{}
\begin{tabular}{@{}lrr}
Name & \multicolumn{1}{l}{HJD} & \multicolumn{1}{l}{Radial velocity}  \\
& \multicolumn{1}{l}{-2400000} & (km\,s$^{-1}$) \\
WD\,1310+583 \\
& 50852.7572&     5.9$\pm$   3.3\\
& 50852.7679&     8.8$\pm$   5.2\\
& 50852.7753&     3.1$\pm$   6.0\\
& 50853.6595&    -3.3$\pm$   5.9\\
& 50853.6658&     7.2$\pm$   5.1\\
& 50853.7830&     3.5$\pm$   3.9\\
& 50853.7902&     4.1$\pm$   3.8\\
& 50854.5618&    -1.4$\pm$   4.5\\
& 50854.5691&    -4.7$\pm$   6.5\\
& 50854.6392&     4.1$\pm$   3.8\\
& 50854.6465&     8.7$\pm$   3.9\\
& 50622.4786&     5.6$\pm$   3.1\\
& 50622.4937&     2.8$\pm$   3.8\\
& 50623.4525&     7.7$\pm$   3.9\\
& 50623.4667&     5.1$\pm$   3.7\\
WD\,1327-083 \\
& 50967.9597&    43.9$\pm$   1.5\\
& 50967.9654&    43.2$\pm$   1.0\\
& 50968.0884&    44.5$\pm$   1.2\\
& 50968.8552&    44.7$\pm$   0.2\\
& 50969.8352&    46.0$\pm$   0.9\\
& 50526.1851&    44.0$\pm$   0.8\\
& 50526.1902&    45.6$\pm$   0.8\\
& 50527.2171&    44.6$\pm$   1.1\\
& 50527.2222&    44.7$\pm$   1.1\\
& 50528.2536&    42.9$\pm$   0.3\\
& 50143.2033&    43.3$\pm$   0.4\\
& 50143.2072&    43.2$\pm$   0.3\\
& 50144.1877&    47.8$\pm$   0.3\\
& 50144.1904&    42.7$\pm$   1.5\\
& 50145.1993&    46.2$\pm$   0.4\\
& 50145.2026&    46.4$\pm$   0.3\\
& 50145.2065&    46.5$\pm$   0.3\\
& 50146.1930&    43.2$\pm$   1.1\\
& 50146.1969&    46.0$\pm$   0.2\\
WD\,1353+409 \\
& 49739.7582&     6.7$\pm$   5.4\\
& 49739.7688&     7.6$\pm$   5.1\\
& 49739.7794&    -7.7$\pm$   6.1\\
& 49739.7918&     3.4$\pm$   5.4\\
& 49162.4427&    -2.4$\pm$   4.4\\
& 49162.4602&    -4.0$\pm$   4.7\\
& 49153.4798&     3.7$\pm$   9.9\\
& 49153.5087&    -1.9$\pm$  12.9\\
& 49150.5101&     5.3$\pm$   5.8\\
& 49150.5212&    -0.1$\pm$   5.8\\
& 49150.5349&     0.6$\pm$   5.5\\
& 50860.6645&    -5.2$\pm$   2.7\\
& 50860.6933&    -6.5$\pm$   2.1\\
\\
\\
\\
\\
\\
\\
\\
\\
\\
\\
\\
\\
\\
\end{tabular}
\end{table}
\begin{table}
\contcaption{}
\begin{tabular}{@{}lrr}
Name & \multicolumn{1}{l}{HJD} & \multicolumn{1}{l}{Radial velocity}  \\
& \multicolumn{1}{l}{-2400000} & (km\,s$^{-1}$) \\
WD\,1407-475 \\
& 50526.2013&    53.9$\pm$   2.0\\
& 50526.2159&    57.2$\pm$   2.0\\
& 50527.2527&    50.0$\pm$   3.0\\
& 50527.2407&    52.0$\pm$   2.9\\
& 50143.2263&    28.5$\pm$   2.2\\
& 50143.2439&    33.3$\pm$   2.0\\
& 50144.1538&    31.0$\pm$   2.0\\
& 50144.1658&    30.1$\pm$   2.0\\
& 50144.2366&    33.7$\pm$   1.9\\
& 50144.2486&    33.1$\pm$   1.9\\
& 50145.2174&    33.0$\pm$   3.3\\
& 50145.2317&    39.4$\pm$   3.8\\
& 50146.2179&    32.4$\pm$   1.8\\
& 50146.2284&    28.1$\pm$   3.6\\
WD\,1422+095 \\
& 50143.2637&     4.9$\pm$   2.9\\
& 50144.1992&     2.4$\pm$   2.3\\
& 50144.2111&     0.1$\pm$   2.2\\
& 50145.2508&    -2.8$\pm$   5.0\\
& 50145.2650&   -10.2$\pm$   4.1\\
& 50146.2722&     7.6$\pm$   3.0\\
WD\,1507+220 \\
& 49888.4890&   -53.3$\pm$   4.2\\
& 49888.5122&   -50.4$\pm$   4.0\\
& 49889.4840&   -52.0$\pm$   3.7\\
& 49889.5027&   -47.6$\pm$   3.7\\
& 49891.4837&   -48.1$\pm$   3.9\\
& 49891.5024&   -45.0$\pm$   3.8\\
& 49892.4597&   -54.6$\pm$   4.5\\
& 49892.4781&   -55.8$\pm$   4.4\\
& 49893.4511&   -54.1$\pm$   6.1\\
& 49893.4694&   -51.6$\pm$   5.9\\
WD\,1507-105 \\
& 51268.2383&   -14.1$\pm$   4.5\\
& 51268.2525&   -13.6$\pm$   6.6\\
WD\,1614+136 \\
& 49155.6365&     2.0$\pm$   4.8\\
& 49155.6561&     4.3$\pm$   7.8\\
& 49162.4850&     7.3$\pm$   4.8\\
& 49162.4963&    17.9$\pm$   4.6\\
& 49162.5935&    11.7$\pm$   4.5\\
& 49162.6049&     0.5$\pm$   4.7\\
& 49153.5447&    17.9$\pm$  12.3\\
& 49153.5635&    -1.5$\pm$  10.3\\
& 49153.5797&    35.5$\pm$  15.3\\
& 49150.5559&    -4.9$\pm$   4.0\\
& 49150.5669&    10.8$\pm$   4.1\\
& 50860.7205&     3.8$\pm$   1.8\\
& 50860.7494&     4.4$\pm$   2.5\\
& 50860.7730&     6.2$\pm$   3.7\\
& 50969.0777&     4.6$\pm$   4.5\\
WD\,1615-157 \\
& 50528.2252&    10.8$\pm$   7.3\\
& 50528.2361&    14.2$\pm$   6.0\\
\\
\\
\\
\\
\\
\\
\\
\\
\\
\end{tabular}
\end{table}
\begin{table}
\contcaption{}
\begin{tabular}{@{}lrr}
Name & \multicolumn{1}{l}{HJD} & \multicolumn{1}{l}{Radial velocity}  \\
& \multicolumn{1}{l}{-2400000} & (km\,s$^{-1}$) \\
WD\,1620-391 \\
& 50676.8555&    45.3$\pm$   0.9\\
& 50676.8593&    46.6$\pm$   0.9\\
& 50676.8631&    47.0$\pm$   0.9\\
& 50677.0418&    44.9$\pm$   2.0\\
& 50677.0435&    44.6$\pm$   2.1\\
& 50677.8541&    47.6$\pm$   1.1\\
& 50968.0006&    47.1$\pm$   1.5\\
& 50528.2034&    48.5$\pm$   1.4\\
& 50528.2115&    47.6$\pm$   1.6\\
& 50528.2972&    50.8$\pm$   1.1\\
& 50528.3052&    51.1$\pm$   1.1\\
WD\,1637+335 \\
& 51071.3364&    29.6$\pm$   4.6\\
& 51071.3505&    34.4$\pm$   3.4\\
& 51071.4081&    20.0$\pm$   3.8\\
& 51072.3472&    25.4$\pm$   3.9\\
& 51072.4038&    26.5$\pm$   3.9\\
WD\,1647+591 \\
& 51067.3415&    48.4$\pm$   3.7\\
& 51067.3460&    40.4$\pm$   3.8\\
& 51067.4252&    41.4$\pm$   3.9\\
& 51067.4289&    41.6$\pm$   3.8\\
& 51068.3308&    42.0$\pm$   6.2\\
& 51068.3345&    40.2$\pm$   5.1\\
& 51068.3382&    43.8$\pm$   4.9\\
& 51068.4093&    40.2$\pm$   3.5\\
& 51069.3372&    42.2$\pm$  10.5\\
& 51069.3483&    27.0$\pm$  22.4\\
& 51069.4029&    39.2$\pm$   5.9\\
& 51071.3235&    61.4$\pm$  38.4\\
& 51071.3260&    30.0$\pm$   6.8\\
WD\,1655+215 \\
& 51067.3588&    43.1$\pm$   3.2\\
& 51067.4136&    37.9$\pm$   3.3\\
& 51068.3483&    42.6$\pm$   4.7\\
& 51068.3589&    32.1$\pm$   4.6\\
& 51072.4521&    50.9$\pm$   8.1\\
WD\,1911+135 \\
& 49888.5600&    23.6$\pm$   3.3\\
& 49888.5684&    16.8$\pm$   3.2\\
& 49889.5480&    26.8$\pm$   3.1\\
& 49889.5564&    23.3$\pm$   3.0\\
& 49891.5509&    23.1$\pm$   4.1\\
& 49891.5589&    20.5$\pm$   3.9\\
& 49892.5223&    18.6$\pm$   3.8\\
& 49892.5303&    15.3$\pm$   3.7\\
& 49893.5143&    19.0$\pm$   5.2\\
& 49893.5234&    12.3$\pm$   5.2\\
WD\,1943+163 \\
& 49888.5912&    36.3$\pm$   3.7\\
& 49888.6003&    39.0$\pm$   3.9\\
& 49889.5695&    33.6$\pm$   4.0\\
& 49889.5778&    35.8$\pm$   4.4\\
& 49891.5725&    42.8$\pm$   5.0\\
& 49891.5804&    40.0$\pm$   4.9\\
& 49892.5447&    30.5$\pm$   4.6\\
& 49892.5527&    34.8$\pm$   4.8\\
& 49893.5403&    40.1$\pm$   6.9\\
& 49893.5494&    40.7$\pm$   8.2\\
& 50969.1790&    34.6$\pm$   4.4\\
& 50969.1863&    35.0$\pm$   4.3\\
& 50970.1076&    35.6$\pm$   2.5\\
& 50970.2356&    36.8$\pm$   2.2\\
\end{tabular}
\end{table}
\begin{table}
\contcaption{}
\begin{tabular}{@{}lrr}
Name & \multicolumn{1}{l}{HJD} & \multicolumn{1}{l}{Radial velocity}  \\
& \multicolumn{1}{l}{-2400000} & (km\,s$^{-1}$) \\
WD\,2058+506 \\
& 51067.5098&    12.5$\pm$   6.0\\
& 51067.5437&    10.4$\pm$   6.0\\
& 51067.5578&    18.4$\pm$   6.7\\
& 51067.6074&    10.0$\pm$  10.5\\
& 51068.5067&    11.0$\pm$  15.6\\
& 51068.6129&    13.1$\pm$   6.0\\
& 51068.6270&     4.8$\pm$   5.2\\
& 51070.5105&    -6.3$\pm$   7.0\\
& 51070.5574&     6.3$\pm$   4.2\\
& 51071.4598&    -2.1$\pm$   5.0\\
& 51071.4704&     5.8$\pm$   4.6\\
& 51071.5634&     9.8$\pm$   5.1\\
& 51071.5741&    12.6$\pm$   4.9\\
& 51071.6386&     8.0$\pm$   4.6\\
& 51072.4717&    22.5$\pm$   7.5\\
WD\,2111+261 \\
& 50759.3525&     7.6$\pm$   4.8\\
& 50759.3649&     1.9$\pm$   4.1\\
& 50760.3017&    -3.4$\pm$   3.3\\
& 50760.3124&    -3.3$\pm$   3.4\\
& 50760.3707&    -2.4$\pm$   2.9\\
& 50760.3814&    -2.1$\pm$   2.6\\
& 50761.3014&    -7.5$\pm$   2.7\\
& 50761.3121&    -4.4$\pm$   2.8\\
& 50761.3627&     1.1$\pm$   2.6\\
& 50761.3734&    -1.9$\pm$   2.6\\
& 50762.3365&    -7.0$\pm$   2.7\\
& 50762.3471&     0.5$\pm$   2.4\\
WD\,2117+539 \\
& 49888.6092&     4.2$\pm$   3.4\\
& 49888.6119&     5.4$\pm$   3.5\\
& 49888.6166&     5.1$\pm$   2.0\\
& 49889.5864&     6.6$\pm$   2.4\\
& 49889.5909&     2.7$\pm$   2.5\\
& 49891.5950&     3.8$\pm$   2.4\\
& 49891.5995&     3.4$\pm$   2.2\\
& 49892.5617&     0.5$\pm$   2.5\\
& 49892.5662&    -3.5$\pm$   2.5\\
& 49893.5595&     3.0$\pm$   3.2\\
& 49893.5646&     5.6$\pm$   3.2\\
WD\,2136+828 \\
& 49888.6248&   -31.8$\pm$   3.2\\
& 49888.6297&   -34.6$\pm$   3.1\\
& 49889.5988&   -35.5$\pm$   3.2\\
& 49889.6033&   -37.7$\pm$   3.3\\
& 49891.6085&   -36.4$\pm$   2.9\\
& 49891.6130&   -33.6$\pm$   2.9\\
& 49892.5743&   -34.7$\pm$   3.7\\
& 49892.5788&   -41.2$\pm$   3.6\\
& 49893.5734&   -38.1$\pm$   5.0\\
& 49893.5779&   -35.5$\pm$   5.6\\
WD\,2151-015 \\
& 51068.5265&    44.4$\pm$   9.6\\
& 51068.5265&    46.4$\pm$   7.7\\
& 51069.5395&    49.3$\pm$   9.3\\
& 51070.5434&    31.6$\pm$   5.6\\
& 51071.4883&    39.4$\pm$   6.1\\
& 51072.4986&    25.0$\pm$   7.8\\
& 50675.9728&    39.4$\pm$   3.9\\
& 50676.1780&    43.9$\pm$   3.4\\
& 50676.1888&    38.4$\pm$   3.6\\
& 50677.2607&    34.6$\pm$  10.2\\
& 50968.2692&    50.4$\pm$   4.3\\
\end{tabular}
\end{table}
\begin{table}
\contcaption{}
\begin{tabular}{@{}lrr}
Name & \multicolumn{1}{l}{HJD} & \multicolumn{1}{l}{Radial velocity}  \\
& \multicolumn{1}{l}{-2400000} & (km\,s$^{-1}$) \\
WD\,2151-015B \\
& 50676.1780&    -5.7$\pm$  10.2\\
& 50676.1888&    -6.1$\pm$  12.8\\
& 50677.2607&     7.5$\pm$   2.5\\
& 50968.2692&    33.3$\pm$  16.8\\
WD\,2226+061 \\
& 49888.6498&    43.9$\pm$   4.2\\
& 49888.6654&    42.7$\pm$   4.0\\
& 49889.6551&    37.4$\pm$   4.0\\
& 49889.6706&    38.7$\pm$   4.3\\
& 49891.6345&    42.8$\pm$   4.4\\
& 49891.6494&    43.8$\pm$   4.5\\
& 49892.5989&    40.9$\pm$   5.4\\
& 49892.6138&    34.8$\pm$   5.1\\
& 49893.5987&    32.2$\pm$   8.0\\
& 49893.6137&    49.5$\pm$   8.3\\
WD\,2341+322 \\
& 51067.6780&    -1.1$\pm$   4.2\\
& 51067.6852&    10.1$\pm$  10.6\\
& 51068.5421&     9.3$\pm$   2.6\\
& 51068.5528&     8.7$\pm$   2.0\\
& 51069.5867&    -4.3$\pm$  10.2\\
\end{tabular}
\end{table}

\section*{Acknowledgements}
 PFLM was supported by a PPARC post-doctoral grant. CM was supported by a
PPARC post-graduate studentship. The William Herschel Telescope and the Isaac
Newton Telescope are operated on the island of La Palma by the Isaac Newton
Group in the Spanish Observatorio del Roque de los Muchachos of the Instituto
de Astrofisica de Canarias.

\label{lastpage} 
\end{document}